\documentclass{article}
\usepackage{amsfonts}
\usepackage{amssymb}
\usepackage{amsmath}
\usepackage{amsthm}
\usepackage{graphicx}
\usepackage[english]{babel}
\begin{document}

\title{\bf Non-Gaussian Probability Distribution\\ for the CMB Angular Power Spectra?}

\author{{\bf Alexey Golovnev}\\
{\small {\it Arnold Sommerfeld Center for Theoretical Physics, Department f\"{u}r Physik,}}\\
{\small \it Ludwig Maximilians Universit\"{a}t, Theresienstr. 37, D-80333, Munich, Germany}\\
{\small Alexey.Golovnev@physik.uni-muenchen.de}}
\date{}

\maketitle

\begin{abstract}

This is my contribution to Proceedings of the International Workshop on Cosmic Structure and Evolution,
September 23-25, 2009, Bielefeld , Germany.
In my talk I presented some non-Gaussian features of the foreground reduced WMAP five
year full sky temperature maps, which were
recently reported in the Ref. \cite{Vitaly}. And in these notes I first discuss the statistics behind this
analysis in some detail. Then I describe invaluable insights which I got from discussions after my talk
on the Workshop. And finally I explain why, in my current opinion, the signal detected in 
the Ref. \cite{Vitaly} can hardly have something to do with cosmological perturbations, but rather it
presents a fancy measurement of the Milky Way angular width in the microwave frequency range.

\end{abstract}

\vspace{1cm}

\section{Introduction}
Nowadays we witness a great progress in both theoretical and observational cosmology which makes our
demands and expectations ever higher and turns us to discussing more and more subtle properties of the
available data. One of such popular topics is the quest for primordial non-Gaussianities in the
spectrum of the CMB radiation. Not really expected to be detectable for the simplest models of inflation,
these small departures from the purely Gaussian signal would help to distinguish between more
elaborate inflationary scenarios and would probably provide us with some new insights into
the wonderful realm of the very early Universe.

The approach I discuss is based on a very simple idea. Assume that the Universe is statistically
isotropic, and all the temperature fluctuations in the CMB radiation are of statistical nature. Then we
decompose the fluctuations, as usual, into the spherical harmonics denoting the coefficients by $a_{l,m}$ and get
$$\langle a_{l,m}a_{l^{\prime},m^{\prime}}^{*}\rangle=\langle a_{l,m}a_{l^{\prime},-m^{\prime}}\rangle
=C_l\delta_{l,l^{\prime}}\delta_{m,m^{\prime}}.$$
Moreover, $a_{l,m}$'s with a fixed value of $l$ but different values of $m$ can
be thought of as different realizations of one and the same random variable. So that one can naturally
ask a question about the shape of the distribution. It can be answered by many methods, from plotting
a histogram of the sample to estimating the higher moments of the distribution.
A Gaussian distribution is completely determined by two parameters, its mean and its variance.
For fluctuations the mean is taken to be $0$, and the only parameter left is the variance, $C_l$.
(Of course, it implies rescaled $\chi^2$-distributions for quadratic in $a$ quantities.)
If the random variable is known (or assumed) to be Gaussian, this single parameter can, in principle, be
extracted from any part of the probability distribution. The idea of the Ref. \cite{Vitaly} is
to take only the tails, i.e. to deduce the variance $C_l$ once more from only the
distribution of large coefficients, $|a_{l,m}|^2>C_l$, and then to compare this result with
the original one. Up to statistical variations in the number of data points in the tails, this
corresponds to using the order statistics with somewhat more points than in top and bottom sextiles but with
less points than in two marginal quintiles.

Applied to the full sky foreground reduced maps, this method gives a well-pronounced peak
in the difference between the two estimates for the variance. The peak is located in
the range of $l\approx 45 \pm 15$. The fluctuations outside of the peak are also much larger
than would be expected statistically. This is, of course, due to remaining foregrounds contamination, and the only
reason to take the peak seriously was that it is a few more times larger than the other fluctuations \cite{Vitaly}
and it looks more or less the same in different frequency bands (up to the different 
overall level of noise). And this was also my conclusion that it should have something to do
with cosmology. I explain the relevant statistics in Sections 2 and 3. And for the graphical
presentation of results, I refer the reader to \cite{Vitaly}.
However, at the Workshop I have learned from Pavel Naselsky that the multipole coefficients with even
values of $l+m$ have the worst contamination from Galaxy, see below. In Section 4 I discuss the
data analysis with separation of $(l+m)$-even and $(l+m)$-odd harmonics, and show that
the initial assumption of having different realizations of one random variable is heavily disproved due
to Galactic signal which invalidates the claim for cosmological non-Gaussianities.
My current conclusion
presented in the Section 5
is that the effects observed in \cite{Vitaly} refer to the structure of Galactic noise, and
not to properties of primordial fluctuations. Due to this reason I cancelled my
authorship for the second version of that article. (I was a co-author for the first one).
And I would like to note here that all the computer work for the article \cite{Vitaly}
was done by my former co-author Vitaly Vanchurin and, needless to say, if there
appears to be something primordial about this peak then the whole success should be
attributed solely to him and his enthusiasm.
An interested reader may also want to consult with
the original reference \cite{Vitaly} for the opinion opposite to mine.

\section{The method and possible variations}

The original approach was to consider $a_{l.m}$'s with $m>0$ and fixed
$l$ as a sample of $l$ observed values of a complex random variable with the Gaussian probability
density of variance $\sigma^2=C_l$:
$$p(z)dzdz^{*}=\frac{1}{\pi\sigma^2}\exp\left(-\frac{zz^{*}}{\sigma^2}\right)dzdz^{*}.$$
Then a function $f_1$, defined by
\begin{equation}
\label{1}
f_1(z)=\frac{e}{2}zz^{*}\cdot{\rm \theta}(zz^{*}-\sigma^2)
\end{equation}
where ${\rm \theta}$ is the step function, has an expectation value equal to the variance. Indeed,
$$\langle f_1\rangle=\frac{e}{2\pi\sigma^2}\int zz^{*}{\rm\theta}(zz^{*}-\sigma^2)\cdot 
\exp\left(-\frac{zz^{*}}{\sigma^2}\right)dzdz^{*}=
\left.\frac{e\sigma^2}{2}\int_{1}^{\infty}ye^{-y}dy=\frac{e\sigma^2}{2}(y+1)e^{-y}\right|_{y=1}=\sigma^2.$$
Statistically, given a sample of $N$ data points (in our case $N=l$), one evaluates the
quantity $${\mathfrak f}_1=\frac{e}{2N}\sum_{i=1}^N z_iz_i^{*}\cdot{\rm \theta}(z_iz_i^{*}-\sigma^2)$$ and
compares it to $\sigma^2$.
Note that in this Section I ignore the fact that we can do nothing but use the sample variance
in (\ref{1}). I'll come to it later.

In order to have more data points we need to resort to real variables because $|a_{l,m}|=|a_{l,-m}|$.
In this case I would assume that there are $N=2l+1$ observations, namely $|a_{l,0}|$, $\sqrt{2}|\Re a_{l,m}|$ and
$\sqrt{2}|\Im a_{l,m}|$ for $m>0$, of a real random variable, in which case the probability density
is
$$p(x)dx=\frac{1}{\sqrt{2\pi}\sigma}\exp\left(-\frac{x^2}{2\sigma^2}\right)dx,$$
and the same analysis carries over for another observable
\begin{equation}
\label{2}
f_2(x)=\frac{1}{1-{\rm erf}\left(\sqrt{\frac12}\right)+\sqrt{\frac{2}{\pi e}}}\cdot x^2{\rm\theta}(x^2-\sigma^2)
\end{equation}
with the error function defined by
${\rm erf}(t)=\frac{2}{\sqrt{\pi}}\int_0^t e^{-x^2}dx$. One can prove prove that
$\langle f_2\rangle=\sigma^2$ using the following simple analytic trick:
$$\frac{4\sigma^2}{\sqrt{\pi}}\int_{\sqrt\frac12}^{\infty}ye^{-y^2}dy=
-\frac{4\sigma^2}{\sqrt{\pi}}\frac{d}{d\alpha}
\left.\left(\int_{\sqrt\frac12}^{\infty}e^{-\alpha y^2}dy\right)\right|_{\alpha=1}=
-\frac{4\sigma^2}{\sqrt{\pi}}\frac{d}{d\alpha}
\left.\left(\frac{1}{\sqrt{\alpha}}\int_{\sqrt\frac{\alpha}{2}}^{\infty}e^{-t^2}dt\right)\right|_{\alpha=1}.$$

All of this can be done in a variety of ways. For example, one can work with order statistics which
I already mentioned in the Introduction. Another possible idea is to average only over $N^{*}$ data points
which are larger than the standard deviation without summing the zeros for excluded entries as
it should be done according to (\ref{1}) or (\ref{2}). The mean number of remaining observations is given by
$\langle N^{*}\rangle=N\int\theta(|x|^2-\sigma^2)p(x)dx$.

In the complex case we have $\langle N^{*} \rangle=\frac{N}{e}$ and transform ${\mathfrak f}_1$
into a new estimator
\begin{equation}
\label{3}
{\mathfrak f}_3=\frac12\cdot
\frac{\sum z_iz_i^{*}\cdot{\rm \theta}(z_iz_i^{*}-\sigma^2)}{\sum {\rm \theta}(z_iz_i^{*}-\sigma^2)}
\end{equation}
if $\exists i:\ z_iz_i^{*}\geqslant\sigma^2$ and ${\mathfrak f}_3\equiv\sigma^2$ otherwise. The latter
case has only a tiny probability with a God-given variance $\sigma^2$, and it is absolutely impossible
if the sample variance is used. Surprisingly, $\langle {\mathfrak f}_3\rangle=\sigma^2$ with no bias.
A simple way to check it is to divide the integration domain into $2^N$ parts with definite signs of 
all $t_i-1$ where $t_i\equiv\frac{z_iz_i^{*}}{\sigma^2}$. The integral under consideration is a symmetric 
function of the variables $t_i$, therefore for each $k$, $0\leqslant k \leqslant N$, one can consider
$\frac{N!}{k!(N-k)!}$ indentical integrals with $t_i\leqslant 1$ for $i\leqslant k$ and
$t_j\geqslant 1$ for $j\geqslant k+1$:
\begin{multline*}
\langle {\mathfrak f}_3\rangle=\int {\mathfrak f}_3\cdot \prod_{i=1}^{N}p(z_i)dz_idz_i^{*}=\\
=\sigma^2\left(\int_0^1 e^{-t}dt\right)^N+\frac{\sigma^2}{2}\sum_{k=1}^{N}\frac{N!}{k!(N-k)!}
\prod_{j=k+1}^{N}\int_{0}^{1}dt_j e^{-t_j}\cdot\prod_{i=1}^{k}\int_{1}^{\infty}dt_i e^{-t_i}\cdot
\frac{t_1+t_2+\ldots+t_k}{k}=\\
=\sigma^2 \sum_{k=0}^{N}\frac{N!}{k!(N-k)!} \left(1-\frac{1}{e}\right)^{N-k}\frac{1}{e^k}=\sigma^2.
\end{multline*}
The product symbols denote here the products of $\int dte^{-t}$ operators, and not of the
integrands after the dot.
Note also that if we were to define ${\mathfrak f}_3\equiv 0$ when all the observations are below the
standard deviation, then a tiny bias of $1$ part in $e^N$ would have been there.

Finally, for real variables $\langle N^{*} \rangle=\left(1-{\rm erf}\left(\sqrt{\frac{1}{2}}\right)\right)\cdot N$,
and the corresponding estimator is:
\begin{equation}
\label{4}
{\mathfrak f}_4=\frac{1-{\rm erf}\left(\sqrt{\frac{1}{2}}\right)}{1-{\rm erf}\left(\sqrt{\frac12}\right)+\sqrt{\frac{2}{\pi e}}}\cdot
\frac{\sum x_i^2\cdot{\rm \theta}(x_i^2-\sigma^2)}{\sum {\rm \theta}(x_i^2-\sigma^2)}
\end{equation}
and  ${\mathfrak f}_4\equiv\sigma^2$ if $\forall i:\ x_i^2<\sigma^2$. The proof that
$\langle{\mathfrak f}_4\rangle =\sigma^2$ is exactly the same as for ${\mathfrak f}_3$ but
with binomial series for $\left(\left(1-{\rm erf}\left(\sqrt{\frac{1}{2}}\right)\right)+{\rm erf}\left(\sqrt{\frac{1}{2}}\right)\right)^N$.

\section{A statistical interlude}
Up to this point I was using the variance in all estimators as if it were known exactly.
In reality the sample variance is used, of course. Generically, it should induce
some bias. For example, with $N$ observations of a random variable $x$ with the mean value
$\mu$ one can estimate the variance as  $\frac{\sum (x_i-\mu)^2}{N}$. It is well-known that
if the sample average is used for the mean, then this estimator is only asymptotically unbiased,
that is it has a non-zero bias which tends to zero when $N\to\infty$. An unbiased estimator
is $\frac{\sum (x_i-\mu)^2}{N-1}$. It diverges if $N=1$ which makes a good sense since nobody
can ever measure two parameters with a one simple observation. Otherwise, one combination of
the data is used to define the mean while the others give information about the random deviations.

I will illustrate the bias only for the estimator ${\mathfrak f}_1$, in which case we have
$$\tilde{\mathfrak f}_1=\frac{e}{2N}\sum_{i=1}^N z_iz_i^{*}\cdot{\rm \theta}\left(z_iz_i^{*}-\frac{1}{N}\sum_{i=1}^N z_iz_i^{*}\right).$$
With the variables $t_i\equiv\frac{z_iz_i^{*}}{\sigma^2}$ one gets
$$\langle \tilde{\mathfrak f}_1\rangle=\frac{e\sigma^2}{2N}\prod_{i=1}^N\int_0^{\infty}dt_ie^{-t_i}\cdot
\sum_{j=1}^N t_j{\rm \theta}\left(t_j-\frac{1}{N}\sum_{k=1}^N t_k\right)=
\frac{e\sigma^2}{2}\prod_{i=1}^N\int_0^{\infty}dt_ie^{-t_i}\cdot
t_N{\rm \theta}\left(t_N-\frac{1}{N}\sum_{j=1}^N t_j\right).$$
In the last equality I have used the fact that everything is symmetric with respect to
permutations of $t_i$ variables. Now I introduce a new variable $v=\sum_{i=1}^N t_i$ instead of
$t_N$ and obtain the final result:
$$\langle\tilde {\mathfrak f}_1\rangle=\frac{e\sigma^2}{2}\prod_{i=1}^{N-1}\int_0^{\infty}dt_i\cdot
\int dv\left(v-\sum_{k=1}^{N-1} t_k\right) {\rm \theta}\left(v-\frac{N}{N-1}\sum_{j=1}^{N-1} t_j\right)=
e\sigma^2\left(1-\frac{1}{N}\right)^N.$$
It is clear that the estimator is only asymptotically unbiased, $\langle \tilde{\mathfrak f}_1\rangle=
\sigma^2\left(1+\frac{1}{2N}+{\mathcal O}\left(\frac{1}{N^2}\right)\right)$. In any case, we are not
expected to do better. There are many other difficulties in extracting the CMB data
from observations.

Of course, one can also estimate the standard deviations for the estimated quantities. 
But note that in our case the only
meaning of it is the minimal level of fluctuations which should be there if the origin of the signal is
stochastic. The actual fluctuations are higher. Nevertheless, the standard 
deviation of ${\mathfrak f}_1$ 
can be found as $\sqrt{\langle{\mathfrak f}_1^2\rangle-
\langle{\mathfrak f}_1\rangle^2}$, and one can easily check that, as usual, it reduces
to $\frac{1}{\sqrt{N}}\sqrt{\langle f_1^2\rangle-
\langle f_1\rangle^2}=\frac{\sqrt{5e-4}}{2\sqrt{N}}\sigma^2\approx 1.55\frac{\sigma^2}{\sqrt{N}}$.
But actually, we are interested in the fluctuations of the difference between the two estimations
of the variance. This difference fluctuates less than the individual terms. Indeed,
$$\left\langle \left(\frac{e}{2}zz^{*}{\rm \theta}(zz^{*}-\sigma^2)-zz^{*}\right)^2\right\rangle=
\sigma^4\left(\frac{e^2}{4}-e\right)\int_{1}^{\infty}t^2e^{-t}dt+\sigma^4\int_{0}^{\infty}t^2e^{-t}dt=
\left(\frac{5e}{4}-3\right)\sigma^4.$$ It gives the deviation $\approx 0.63\frac{\sigma^2}{\sqrt{N}}$
which should be approximately correct because the random excursions of $\sigma$
are much less than those of the individual observations.

For the estimators (\ref{3}) and (\ref{4}) the standard deviations are not of the form
$\frac{something\ simple}{\sqrt{N}}$. One can find them exploiting the above trick with the 
integration domains, this time for the integrand $\frac{1}{k^2}\left(\sum_{i=1}^k t_i\right)^2$.
The result would be
$$\langle {\mathfrak f}^2_3 \rangle=\frac{\sigma^4}{4} \sum_{k=1}^{N}\frac{N!}{k!(N-k)!} \left(1-\frac{1}{e}\right)^{N-k}\frac{1}{e^k}
\cdot\frac{4k^2+k}{k^2}=\sigma^4+
\frac{\sigma^4}{4} \sum_{k=1}^{N}\frac{N!}{k!(N-k)!} \left(1-\frac{1}{e}\right)^{N-k}\frac{1}{e^k}\cdot\frac{1}{k}.$$
The first term is just $\langle {\mathfrak f}_3 \rangle^2$ while the second one gives
the variance. Asymptotically, only large values of $k$ do matter, and we can substitute
$k$ by $k+1$. Up to the factor of $\frac{e}{N+1}$ it gives the binomial formula again (neglecting
the contribution of the first few terms). Hence, the standard deviation is given by
$\approx\frac{\sqrt{e}\sigma^2}{2\sqrt{N}}\approx 0.8\frac{\sigma^2}{\sqrt{N}}$. A little bit
more tedious calculation shows that the standard deviation of the difference between
$\langle{\mathfrak f}_3\rangle$ and the sample variance has the same asymptotic behaviour.

Ideally, one should also take the effects of using the sample variance in the step functions into account.
But it is not of my concern now. For clean maps the main task would have been to test the hypothesis
that all the deviations from zero difference between two $C_l$ estimations are purely statistical. However,
we use the noisy maps. And the signal considerably deviates from Gaussianity anyway. The curious result was only
about the large peak at $l\approx 45 \pm 15$ which was argued to have a cosmological origin \cite{Vitaly}.

\section{The structure of Galactic contamination and the data}
My talk at the Workshop was followed by a very interesting discussion, and I learned from Pavel Naselsky
that the $a_{l,m}$-coefficients with even values of $l+m$ are more contaminated by the Galaxy \cite{Pavel}.
The reason is very simple to understand examining the Rodrigues formula for the associated
Legendre polynomials on the interval $z\in [-1,1]$,
$$P_l^m(z)=const\cdot \left(1-z^2\right)^{\frac{m}{2}}\frac{d^{l+m}}{dz^{l+m}}\left(z^2-1\right)^l.$$
The Galactic plane corresponds to $z=0$. Therefore, $(l+m)$-odd polynomials are antisymmetric
under reflection in Galactic plane ($z\to -z$) and have roots at $z=0$, while the  $(l+m)$-even
ones have local extrema at the same place. Note also that $l\sim 45$ corresponds to the scale
of several angular degrees which nicely matchs with the apparent width of the Milky Way and with the width
of the red central stripe on the pictures of the WMAP results. One has to analyse these harmonics
separately. And as reported in the second version of \cite{Vitaly}, the $(l+m)$-even coefficients
reproduce mostly the same shape of the super-Gaussian peak, while the $(l+m)$-odd ones give a small
sub-Gaussian valley at the same values of $l$. It already shows that different $a_{l,m}$'s for
the same $C_l$ are not at all equivalent as they were assumed for the purposes of this analysis.

I also handled the W band data manually in order to gain a better perception of the numbers.
Examining the data, one can see the super-Gaussian character of $(l+m)$-even observations with almost
a naked eye. There are many points with large values, and many points are considerably smaller
than the standard deviation. Sometimes just a couple of very large data points makes a significant
part of the peak. For example, I found that the imaginary part of $a_{46,46}$ is more than two times
larger than even the largest of other numbers in the group of $C_{46}$.

\begin{figure}
\centerline{\includegraphics[height=.4\textwidth]{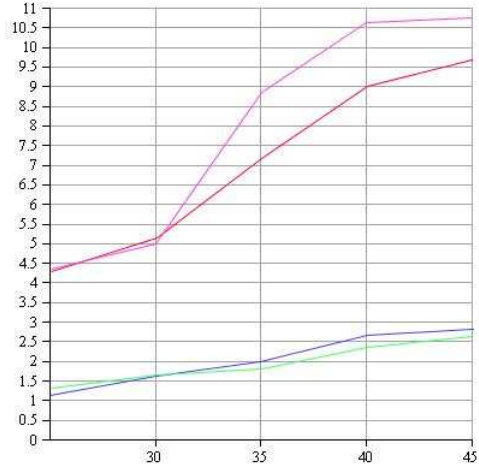}}
\caption{Estimated values of $10^{-3}\cdot\frac{l(l+1)C_l}{2\pi}$ in units of $\mu{\rm K}^{2}$
for the W band. The red line is obtained with all $(l+m)$-even harmonics. For the purple
line only large $(l+m)$-even coefficients are used. The $(l+m)$-odd multipoles give the blue line,
while restricting them to only the large ones results in the green line.}
\label{fig1}
\end{figure}

 On the Fig. 1
I present the values of $\frac{l(l+1)C_l}{2\pi}$ versus $l$. Only the left, ascending part of the
peak is plotted there, as it is less challenging for a manual computation. Unlike in \cite{Vitaly},
I used the estimator (\ref{4}). The data points are binned into the groups of five, that is
the values of the function are given only for values of the argument congruent to zero modulo
five, and each ordinate is an average of five values, those from $l-2$ to $l+2$.
For even harmonics we see the peak, where the red line is the sample variance for all
even coefficients, while the purple line is obtained with the estimator (\ref{4}).
The same is done for odd harmonics (the blue and the green lines), and a small valley
is revealed. I won't bet for the
precise ordinates as I was calculating manually and rounding the data a little bit.
But the general structure is represented correctly and agrees with the claim in the Ref. \cite{Vitaly}. 
We can see that the peak is somewhat reduced compared to what was plotted in \cite{Vitaly} without 
the separation of harmonics, at least in its relative weight, $\frac{\Delta C_l}{C_l}$. And what is more
important, the red line goes several times higher than the blue one. (And even the blue line is some
factor of two higher than the actual primordial radiation \cite{Gary}.) This is true not only in average, but
also for every single $C_l$. It clearly shows that the dominant
signal for these multipoles comes from the Galaxy, and it also heavily disproves the original
hypothesis of having different observations of a one random variable. If $l+m$ is even, 
one also gets the large coefficients with large $m$'s
more frequently than for the middle values of $m$, as was stated in \cite{Vitaly}. This effect is
by far less pronounced than the difference between the red and the blue lines. But on the other hand,
it shows that, even after the separation of different parity harmonics, the signal cannot be
analysed reliably in this way. A considerable part of the initial peak came from the mixing of 
harmonics with essentially different levels of contamination which overweights the central part and
the tails of the distribution. The remaining 
$(l+m)$-even peak may also be the consequence
of a non-uniform contamination, although the Galactic signal by itself is not very Gaussian. It could also
be an interesting problem to compare the results of real and complex random variables analysis.
The direction of zero Galactic longitude points at the Galactic center, and therefore cosine
and sine harmonics might receive different contaminations. 

The peak disappears when we go to larger $l$ and the spherical harmonics
start probing the latitude scales smaller than the width of the bulk of Galactic signal. And therefore
the different parity harmonics become not so different in the contamination level. For example,
$\frac{l(l+1)C_l}{2\pi}$ for $l=100$ estimated with $(l+m)$-even harmonics is only few percent larger than estimated with 
$(l+m)$-odd ones. (It is about $8000 \mu K^2$.) Moreover, $|a_{100,0}|$ is quite large, and $C_{100}$-even
estimated without $m=0$ is a bit smaller than $C_{100}$-odd.

It is hard to infer about the origin of the valley without knowing the detailed structure of the
Galactic noise. It can come from some non-Gaussian properties of the noise. 
It can reflect some shortcomings of the
foreground reduction procedures somehow oversubtracting the super-Gaussian noise from less
affected coefficients. Probably, one could even devise a reasonable mixing of signals which
would mimic a sub-Gaussian distribution, although it may require some bias of the mean values too.

\section{Conclusions}
In these notes I discussed the statistics behind the non-Gaussian anomalies reported in Ref. \cite{Vitaly}. After that
I have shown that the most probable explanation of the signal
refers to geometric properties of the Galactic noise, and not to cosmology. Admittedly, I do not
have a good understanding of the structure of Galactic noise. But at the very least,
the claim for cosmological non-Gaussianities is pretty much premature. (I refer an interested
reader to the work \cite{Vitaly} for a different opinion.) On the other hand, one could probably
use this kind of analysis to extract some information about the structure of the noise.
 
{\bf Acknowledgments.}
I am very grateful to the organizers of the Workshop for the opportunity to participate
in this wonderful event, and especially I wish to thank Dominik Schwarz for invitation and
for his encouragement to write this contribution. I am also very grateful to Pavel Naselsky 
and to other participants for very useful discussions. This work was supported in
part by the Cluster of Excellence
EXC 153 {}``Origin and Structure of the Universe''.

\end{document}